# Model complexity in econometrics – a combinatorial analysis

Vahidin Jeleskovic[1]

ABSTRACT

Regression models and Vector Autoregressive Models (VARs) play crucial roles in econometrics by allowing the analysis of multiple variables simultaneously. Despite their utility, these models face challenges like underfitting and overfitting, especially when determining the optimal model specification, which can lead to significant computational costs. To address these challenges, econometricians often rely on widely adopted model selection criteria such as the Akaike Information Criterion (AIC) and Bayesian Information Criterion (BIC). These criteria help balance model complexity and goodness of fit, aiding in the selection of the most suitable model specification for the given data. Nonetheless, there is a notable gap in existing research concerning the correct specification of these models, particularly in determining the optimal number of states a system can assume. Addressing this gap, we introduce a combinatorial framework designed to calculate the potential number of states in such econometric models. Our approach involves delineating four distinct stages in model development, each offering a range of specifications. This method enables a comprehensive combinatorial calculation of all possible states. The aim of this paper is to highlight this overlooked aspect of model specification and to spark a constructive dialogue within the empirical research community. By doing so, we hope to inspire further research that enhances the precision and applicability of econometric models. A theoretical complexity criterion is necessary to elucidate fundamental limitations and propose new objectives to pursue.

[1] Humboldt-Universität zu Berlin, Spandauer Str. 1, D-10009 Berlin. Email: vahidin.jeleskovic@gmail.com

## 1. Introduction

Econometric models bridge the gap between theoretical models on a specific issue and empirical real-world data, known as the empire. These models facilitate the examination of relationships between one or more dependent variables of interest and explanatory variables, which may include lags of the dependent variables. Central to this process is the assumption of correct model specification. If the econometric model omits relevant explanatory variables, its estimators might be biased, leading to biased or false results and conclusions. Conversely, incorporating too many explanatory variables can inflate the variance of the estimators, reducing the model's precision. Hence, researchers are faced with the trade-offs in econometric modeling between bias and variance. This is especially critical when the dataset is relatively small, leading to a very limited number of degrees of freedom. Thus, if the assumption of a correctly specified model is violated, coefficient estimates, confidence intervals, and hypothesis tests may become misleading and inconsistent, a problem known as specification error. The issue of specification error and its implications were notably discussed in the early publications edited by Kmenta and Remsy (1980).

In the context of model selection, commonly employed metrics include e.g. the adjusted $R^2$, the Akaike Information Criterion (AIC), the Bayesian Information Criterion (BIC), and the Hannan-Quinn Information Criterion (HQC). These criteria are designed to balance model fit with complexity, penalizing models with an excessive number of parameters to prevent overfitting. While these criteria are useful, they do not offer insights into the potential number of model misspecifications. Grasping the underlying optimization challenge, specifically attaining the correct model specification, is crucial for identifying the optimal solution, which in this context is the most suitable model specification. Therefore, accurately determining the number of misspecifications is key to understanding the range of all possible solutions. This understanding provides a comprehensive view of the model specification landscape, allowing for more informed and effective choices in model selection.

The aim of this paper is to illuminate the complexities inherent in optimizing econometric model specifications and to quantify the extent of these challenges. Initially, our focus is usually on a singular econometric model, contextualized around a specific variable of interest. In this context, an econometric model is understood as one where the variable of interest is modeled based on a comprehensive information set. This set includes all pertinent explanatory variables, as well as their lagged counterparts and lags of dependent variables. Subsequently, we broaden our examination to include multiple variables of interest, each with its respective relevant information set.

This extension naturally leads to an exploration of their interrelationships, thus elevating the complexity from a single-model framework to a multi-model paradigm, which is more realistic from the perspective of the real world. This exploration is commonly achieved by applying VAR (Vector Autoregressive) and VARX (Vector Autoregression with Exogenous Variables) models, which additionally considers exogenous variables. models. Our initial focus at first is on the reduced form of the VAR model. In this form, each variable is typically expressed as a function of its own lagged values and those of all other variables in the model, up to a specified maximum lag order. Consequently, the flexible yet optimal specification of a VAR model in its reduced form is crucial for developing and interpreting its structural form. The relationship between these two forms in econometrics is complex, and the quality of the reduced form directly impacts the structural form (citation). The reduced form of a VAR model provides the empirical foundation from which the structural form is derived (citation). The structural form is based on theoretical considerations and hypotheses about causal relationships between variables. Therefore, a precise and well-specified reduced form is critical for arriving at valid and meaningful structural models. However, this approach doesn't necessarily imply that all lags up to the highest order in the reduced form should always be included for each dependent variable in each equation. For instance, the market's response to certain events might materialize in the next period, or it could take two to three periods. This delayed response is also observable in scenarios like a company adjusting its production levels over time, or the delayed effects of fiscal policies. These varying lags necessitate careful consideration in the model specification to accurately capture the dynamics of the system.

The inclusion of lags is subject to several considerations, primarily the Lag Order Selection. The maximum number of lags included in a VAR model is usually determined based on information criteria such as AIC, BIC, or HQC. In cases where advanced modeling techniques are employed, or when theoretical frameworks indicate distinct dynamics for various variables, researchers may choose lags specific to each variable. This means that within the VAR model, the number and the structure of lags can differ across variables and equations, introducing additional complexity to the model. This principle of variable-specific lags is also applicable in VARX models.

To provide greater specificity and vividly illustrate the issue, we aim to pose a concrete research question regarding the rapid escalation in complexity encountered when tackling real-world econometric problems. For an initial exploration, consider models used in analyzing stock price returns with different set of exogenous variables. Prominent among these are well-known the Capital Asset Pricing Model (CAPM) as well as Fama and French's three-factor and five-factor models. Additionally, other models can be considered for analyzing returns, such as autoregressive (AR) models and ARX models which include further exogenous variables like those based on a company's Key Performance Indicators (KPIs). Given this diversity, a meticulous model specification is essential. It should encompass a comprehensive information set that includes all relevant variables to effectively model stock price returns. This approach ensures a thorough and accurate analysis, capturing the multifaceted nature of variables of interest.

Other economic variables have gained a lot of interest as well. The gross domestic product (GDP) is undeniably a central focus of attention for academics, policymakers, and economists alike, warranting extensive empirical analysis due to its pivotal role in the real economy. However, it's crucial to recognize the intricate interplay between the financial markets and the real economy, suggesting the need for a simultaneous examination of GDP alongside financial market variables. This can be effectively accomplished through the application of a VAR (Vector Autoregressive) model, where variables such as stock returns and GDP are modeled based on their historical values, or through a VARX model that incorporates further exogenous variables into the analysis.

Furthermore, expanding our analysis to include additional factors (or variables) which interacts with GDP and financial markets could be the foreign direct investments (FDI), broadens our scope significantly. In this expanded scenario, we are dealing with a system that models three dependent variables, namely GDP, FDI, and financial market indicators. The potential to further augment this system is considerable, for example by incorporating variables of interest like unemployment rate, health industry, trade balance, climate dynamic etc. Each of these variables of interest brings its unique information set, yet all are interconnected within the system. Each model not only leverages its specific information set but may also be influenced by others. This approach allows for a more nuanced exploration of model specification complexity in econometrics.

This expansion leads to a pivotal question: In such an intricate system, how many model specifications are possible, and which is the most effective? Addressing this question is essential for comprehending the challenges in optimizing model specifications within a network of interconnected economic variables. The optimal system specification can only be determined by considering all possible states of the system. This comprehensive exploration is crucial to fully understanding the complexities involved in optimizing model specifications in econometrics.

The paper presents, in the next section, the combinatorial approach for calculating all possible model specifications and is fully flexible in this sense. The discussion is then provided in the last section.

## 2. Combinatorial approach for calculation the number of model specification

To illustrate and quantify the complexity inherent in this comprehensive model, we utilize a combinatorial approach. This model's optimization challenge lies in accurately specifying various interacting variables of interest. For each variable, there is an associated information set of explanatory variables which includes the lags of all dependent variables as well as some exogenous variables in the case of VARX. Within this set, we must select either a specific subset or all explanatory variables. These variables may hold key information that impacts every dependent variable in the model, potentially leading to crossover effects. This approach highlights the intricate interplay of variables and the complexity of achieving optimal model specification. To calculate the number of potential states in which a VAR or VARX model can exist, a new framework is required. This framework aims to establish a general concept for this class of models, where each specification represents a selection of elements from finite sets, determined by the researcher conducting the specification. This approach allows for a systematic and comprehensive enumeration of all possible model configurations, facilitating a deeper understanding of the model's potential states and the implications of different specification choices.

In the framework we propose, both VAR and VARX models are conceptualized as embodying multiple key aspects: multi-objective, multi-disciplinary, multi-dimensional, and multi-frequent. 'Multi-objectivity' in this context refers to the incorporation of several dependent variables. 'Multi-explanatory' highlights the inclusion of numerous explanatory variables on the right-hand side of the equations for each dependent variable. 'Multi-frequent' denotes the varied frequencies at which these variables can be observed or measured. 'Multi-disciplinarity' is crucial in the concept for our approach and is needed to be explained in more details.

In our approach, we introduce a simplifying assumption by uniquely associating each explanatory variable with a specific subject area or dataset, which we define as a 'discipline.' This 'discipline' is considered both distinct and constant. Incorporating lagged dependent variables into our models highlights cross-disciplinary influences, emphasizing the complex interconnections and dependencies inherent in this elaborate modeling framework. Within this framework, the aggregated set of lagged variables forms a unique 'discipline' for each equation. Specifically, in the VAR model with k dependent variables, we encounter a singular discipline that encapsulates all lagged dependent variables as explanatory factors. Conversely, the VARX model allocates to each dependent variable its own discipline, consisting of all directly related explanatory variables, ensuring that these disciplines are mutually exclusive. Furthermore, the interactive dynamics of the equation system necessitate the inclusion of all dependent variables' disciplines in each equation. Therefore, with k dependent variables, our framework introduces m = k+1 distinct and non-overlapping disciplines, laying the groundwork for our 'multi-disciplinarity' concept. Notably, despite each discipline's uniqueness, they collectively exert direct influences on every dependent variable. By delineating disciplines to prevent variable overlap, we refine our combinatorial strategy, significantly enhancing its efficiency and effectiveness in navigating the complexities of econometric modeling.

Formally, this means that both VAR and VARX models are composed of k target or dependent variables, denoted as $y_i$, where $i = 1, \dots, k$. For each dependent variable $y_i$, $m_i$ disciplines $D_{i,j}$, where $j = 1, \dots, m_i$, are assigned to ensure complete flexibility in model specification. However, we initially postulate that the maximum number of disciplines assignable to each equation is $m_i = k + 1$. This assumption allows for comprehensive coverage and interaction among variables, while also facilitating a structured approach to model complexity and specification.

We designate $D_{i1}$ as the discipline comprising all lagged dependent variables. If the disciplines $D_{ij}$ for j = 2,…, k+1 contain no variables, the model in question is a VAR model. Conversely, if any of these disciplines contain one or more variables, we classify the model as a VARX model. Furthermore, should all lagged dependent variables be considered non-contributing (or zero variables) and at least one exogenous variable is introduced as an explanatory variable in each equation for every dependent variable, we then encounter a system of k independent equations. It is important to note that such a scenario, implying complete independence among equations, falls outside the purview of this paper's analysis.

For VAR and VARX models with $k$ dependent variables we define the maximum lag order $p_{e,i}^{max}$ of the dependent variable $e$ in equation i, where $i = 1, \dots, k$, we impose a constraint that each equation for every dependent variable must contain at least one explanatory variable. The intercept is also set in each equation of VAR and VARX whereas the existence of the intercept can be ignored within this combinatorial or approach as it is included in each equation. Under these conditions, the number of possible states for such a model with $k$ dependent variables can be calculated as follows:

$$\boldsymbol{C_{k,D}} = \prod_{i=1}^{k} \sum_{j=1}^{q_i} \binom{q_i}{q_j} \tag{1}$$

$\boldsymbol{q_i}$ considers the number of all explanatory variables in equation $i$ and it is calculated as follows:

$$\boldsymbol{q_i} = \sum_{e=1}^{k} \boldsymbol{p_{e,i}^{max}} + \sum_{e=1}^{k} \boldsymbol{d_{e,i}} \tag{2}$$

where:

$k - the\ number\ of\ dependent\ variables;\ i = 1, \dots, k$

$p_{e,i}^{max} - the\ maximum\ lag\ order\ of\ the\ dependent\ variable\ e\ in\ equation\ i;$
$e\ = 1, \dots , k$

$d_{(e\ ,i)} - the\ number\ of\ exogenous\ variables\ in\ the\ discipline\ D_e\ in\ equation\ i,$

$including\ the\ lags\ of\ the\ exogenous\ variables;\ \ e = 1, \dots , k$

This methodology affords a fully flexible framework for the construction of VAR and VARX models, freeing us from the rigid assumption of a uniform maximum lag order across all dependent variables and equations. Instead, it permits the specification of distinct maximum lag orders for each dependent variable and their respective equations. Additionally, our approach facilitates the determination of a varying number of explanatory variables within each discipline for every dependent variable, enhancing the model's adaptability and precision in capturing economic dynamics. Consequently, our approach can be seamlessly implemented within a software environment, enabling rapid computation of the total number of possible states for a concrete system of equations.

However, for illustrative purposes and to simplify our discussion, we adopt a common econometric practice of assuming a uniform maximum lag order across all equations in both VAR and VARX models. As an example, consider a VAR model featuring three dependent variables ($k = 3$), consistent with our earlier discussion, and assign a maximum of four lags to each as in our previous example above, and a maximum of four lags for each ( $p_{i,l}^{max} = 4, \forall\ i \in \{1,2,3\}$), a scenario typical in quarterly data modeling, such as that of GDP. Under these conditions, we proceed to calculate the total number of possible states as follows:

$$q_i = \sum_{i=1}^{3} 4 + \sum_{i=1}^{3} 0 = 12$$

$$C_3 = \prod_{i=1}^{3} \sum_{j=1}^{12} \binom{12}{q_j} = 68.669.157.375$$

The revelation of approximately 68,669 billion possible states underscores the profound complexity inherent in even the most basic configurations of VAR models, illuminating the substantial intricacies of econometric modeling. This highlights the nuanced depth required to grasp the full spectrum of potential outcomes and interactions within what might initially appear as a straightforward VAR model.

Transitioning to VARX models, this complexity further escalates with the introduction of exogenous variables. To illustrate, consider the scenario where each discipline incorporates two exogenous variables. This adjustment must be applied across all three dependent variables, necessitating a recalculated approach as follows:

$$q_i = \sum_{i=1}^{3} 4 + \sum_{i=1}^{3} 2 = 18$$

$$C_{3,\{2,2,2\}} = \prod_{i=1}^{3} \sum_{j=1}^{18} \binom{18}{q_j} = 18.014.192.351.838.208$$

Hence, the number of all possible states of such VARX model is approximately 18.014 * $10^{15}$. This demonstrates the even greater complexity of a VARX model compared to a VAR model, particularly when accounting for multiple exogenous variables.

Let's assume a scenario and address the computational feasibility of calculating an information criterion (such as AIC or BIC) for every single model specification derived from a time series of length 1000 using Ordinary Least Squares (OLS) estimator on a supercomputer operating at petaFLOP speeds. It would probably still need years to calculate an information criterion for every single model specification in this case.[2]

Yet, it's important to note that the real world is even much more complex than this modeling suggests. Until now, our analysis has operated under the assumption that the world functions on a single time frequency, a significant oversimplification that doesn't accurately represent the complex dynamics of real-world phenomena. It's evident that this narrow perspective fails to encompass the multi-faceted and multi-temporal nature of real-world events, which typically operate across multiple time frequencies. Take, for instance, our model with three dependent variables: GDP, FDI, and stock prices. While analyses can be performed e.g. on a yearly basis, it's reasonable to hypothesize that critical dynamics for these variables also transpire at more granular intervals, such as quarterly or monthly. While the GDP may be measured quarterly, other variables like inflation rate, stock prices and exchange rates are measured on much higher frequencies up to tick data and thus contain much more information. Hence, we must consider a situation where at least one of dependent variables contains more information due to its measurement on higher frequency. Otherwise, if all variables are observed at the one and the same highest frequency, then there is no rational reason why not to provide an econometric analysis at that highest frequency for all dependent variables. Some econometric models have been proposed, e.g. so-called MIDAS by Gysels et. al (2002) and so-called Multifrequency VAR models by Gysels (2015) (see also the references therein) to catch these additional

[2] Practical Limitations would be, even with parallel processing, not just about raw computations (which are massive) but also involves data management, memory access, and the overhead of managing 18*$10^{15}$ distinct computational tasks which include several steps within each of them.

information on higher frequencies and propose econometric modeling for that.[3] When incorporating multiple frequencies $F_i$ of the dependent variable $i$, where again $i = 1, \dots, k$, and maintaining the model specification as previously described, the calculation formula is as follows:

$$\boldsymbol{C_{k,D,f}} = \prod_{i=1}^{k} \sum_{j=1}^{q_{i,f_i}} \binom{q_{i,f_i}}{q_j} \tag{3}$$

where $f_i$ denotes the total number of time points for a specific variable $e$ on the right-hand side of equation $i$ within one period. E.g. for $f_i = 1$ it would be the base frequency like quarterly data and for $f_i = 4$ it would be higher frequency in terms of monthly data and for $f_i = 16$ for weekly frequency in this particular example. That is, it represents the highest sub-frequency within the set of all frequencies $F_i$ for that variable. This is because, if we consider a variable measured monthly, it aligns with quarterly data at the end of March, June, September, and December, respectively. Consequently, the highest frequency inherently encompasses all lower frequencies. Therefore, the number of explanatory variables in the equation $i$ can be determined as follows:

$$\boldsymbol{q_{i,D,f_i}} = \sum_{e=1}^{k} \boldsymbol{f_{e,i}} * \boldsymbol{p_{e,i}^{max}} + \sum_{e=1}^{k} \boldsymbol{f_{e,i}^{d_{e,i}}} * \boldsymbol{d_{e,i}} \tag{4}$$

where

$f_{e,i} - the\ highest\ frequency\ of\ the\ variable\ e\ in\ equation\ i$

$f_{e,i}^{d_{e,i}} - the\ highest\ frequency\ of\ the\ variable\ d_{e,i}\ within\ discipline\ e\ in\ equation\ i$

If we don't consider any sub-frequency, that is there is only the base frequency with $f_{e,i} = 1$ and $f_{e,i}^{d_{e,i}} = 1$ then (3) is equal to (1).

To simplify, let's consider a multi-frequency model for the above example with three dependent variables which are now measured on the annually frequency. We think that the dynamics on quarterly and/or monthly frequencies are different from the base frequency (here yearly) and have significant impacts on the dynamic of dependent variables on the base frequency. For a VAR(4) with 3 dependent variables on the yearly base frequency with the quarterly sub-frequency, we would have following number of states:

[3] We do not consider modeling approaches that impute data points at higher frequencies based on observations at lower frequencies e.g. using state-space models or Bayesian methods. This is because, in the context of our analysis, such approaches practically coincide with the VAR and VARX as discussed in this paper.

$$q_{i,0,4} = \sum_{1}^{3} 4 * 4 = 48$$

and

$$C_{3,0,4} = \prod_{i=1}^{3} \sum_{j=1}^{48} \binom{48}{q_j} \approx 2.230075 * 10^{43}$$

If we consider the monthly sub-frequency in this setting:

$$q_{i,0,12} = \sum_{1}^{3} 12 * 4 = 144$$

and

$$C_{3,12} = \prod_{i=1}^{3} \sum_{j=1}^{144} \binom{144}{q_j} \approx 1.109068 * 10^{130}$$

Thus, the total number of potential states in even a basic multi-frequency VAR (Vector Autoregression) model could exceed the estimated number of atoms in the universe, which ranges from $10^{78}$ to $10^{82}$, or even surpass the Shannon number, which encapsulates the entire spectrum of potential move sequences in chess, estimated between $10^{111}$ to $10^{123}$ when considering both legal and illegal moves.

Now, we expand our theoretical experiment to include a VARX(4) model, which incorporates two exogenous variables from each discipline, also observed at quarterly sub-frequencies:

$$q_{i,\{2(4),2(4),2(4)\},4} = \sum_{1}^{3} 4 * 4 + \sum_{1}^{3} 4 * 2 = 64$$

and

$$C_{3,\{2(4),2(4),2(4)\},4} = \prod_{i=1}^{3} \sum_{j=1}^{64} \binom{64}{q_j} \approx 6.277102 * 10^{57}$$

And finally, when considering the monthly sub-frequencies, we achieve the following number of model specifications:

$$q_{i,\{2(4),2(4),2(4)\},12} = \sum_{1}^{3} 12 * 4 \; + \sum\nolimits_{1}^{3} 12 * 2 = 216$$

and

$$C_{3,\{2(12),2(12),2(12)\},12} = \prod_{i=1}^{3} \sum_{j=1}^{216} \binom{216}{q_j} \approx 1.16798 * \; 10^{195}$$

Therefore, as we observe, the immense figures representing the total number of possible model specifications become almost absurd, even when attempting to employ relatively straightforward models like VAR or VARX. This complexity arises when we consider 3 dependent variables, each with 4 lags, and two independent variables in each discipline, across all 12 sub-frequency points to accommodate the multifrequency nature of real-world phenomena. Yet, our exploration into the intricacies of VAR and VARX models, in terms of the myriad states that must be considered, is far from over. This journey into understanding the comprehensive complexity of these models is ongoing, highlighting the challenge of accurately capturing the dynamic interplay of variables in econometric analysis.

Up to this point, our approach has been based on the assumption on the premise that we possess a clear understanding of the number of dependent variables and their mutual interactions. For instance, we assume that GDP, FDI, and financial market indicators engage in a mutual interaction encapsulated within a system of equations—a notably restrictive presumption. This perspective allows for the possibility that GDP not only correlates with FDI and financial market but also suggests that an array of variables including GDP, FDI, the financial market, along with trade balance, climate dynamics, and more, could all be interlinked. This opens up a broader view of economic interdependencies, challenging us to consider a more complex network of relationships than initially assumed. We must also firmly assume the total number of variables that interact and depend on each other ($K^{max}$). Moreover, we assume that each variable, such as GDP, can be represented by a single equation, allowing us to determine the overall complexity of the econometric model:

$$\boldsymbol{C^{global} = \sum_{I=1}^{K^{max}} C_{I,D_I,f_I}} \tag{5}$$

## 3. Discussion

In the context of VAR and VARX models, the challenge of finding the optimal model specification involves selecting the appropriate variables, determining the correct lag length for each variable, and possibly deciding on the inclusion of exogenous variables and dynamics on sub-frequencies. This process can be viewed through the lens of an optimization problem, where the goal is to maximize some criterion of model fit (e.g., minimizing information criteria like AIC, BIC, etc.) while balancing the complexity of the model. The combinatorial nature of this problem (choosing among a potentially vast number of variable combinations and lag lengths) can lead to an exponential growth in the number of possible model specifications as the number of candidate variables increases. This combinatorial explosion resembles the computational intractability characteristic of NP-hard problems, particularly when considering all possible subsets of variables and all feasible lag configurations.

The concept of NP (nondeterministic polynomial time) problems and the formalization of the class NP were first introduced in the early 1970s. The term "NP" and the concept of NP-completeness were popularized by the work of Stephen Cook and, independently, Leonid Levin.

Stephen Cook's seminal paper, "The Complexity of Theorem-Proving Procedures", was presented at the Third Annual ACM Symposium on Theory of Computing in 1971. In this paper, Cook introduced the concept of an NP-complete problem, using the Boolean satisfiability problem (SAT) as the first example of such a problem. This work laid the foundation for the theory of computational complexity. Leonid Levin, working independently in the Soviet Union, came to similar conclusions and identified a list of six NP-complete problems. His work was published in 1973. These papers laid the groundwork for understanding the complexity of decision problems for which a given solution can be verified in polynomial time (NP).

The concept of NP-hardness, which describes problems that are at least as hard as the hardest problems in NP (but not necessarily in NP themselves), was developed in conjunction with the concept of NP-completeness. The term "NP-hard" was introduced shortly after Cook's presentation, notably by Richard Karp in his 1972 paper "Reducibility Among Combinatorial Problems," where he extended Cook's work by identifying 21 NP-complete problems, effectively showing the widespread impact of NP-completeness across various areas of computer science and mathematics.

Therefore, the concept of NP-hardness emerged essentially simultaneously with or immediately after the formalization of NP problems, as the understanding of the complexity of these problems and their categorization naturally led to the identification of problems that are as difficult as the hardest problems in NP. A subset of NP-hard problems, called NP-complete problems, are those that are both in NP and NP-hard. These problems have the property that a solution can be verified in polynomial time, and they encapsulate the dual hardness of being difficult to solve but easy to check.

The question of whether NP-hard problems can be solved in polynomial time is central to the famous "P vs. NP" problem, one of the seven Millennium Prize Problems. It asks whether every problem whose solution can be verified in polynomial time (NP) can also be solved in polynomial time (P).

While previous research has explored various reduction methods for VAR models, such as those discussed by Brüggemann et al. (2003), the task remains daunting due to the multitude of potential model specifications and the associated computational complexity.

One prominent approach to model reduction is the Least Absolute Shrinkage and Selection Operator (LASSO), which aims to enhance prediction accuracy and interpretability by shrinking less important variables' coefficients towards zero. However, LASSO is not without its limitations, including concerns about biased estimates, handling correlated predictors, and the selection of tuning parameters. Moreover, given the vast number of possible model specifications, there's a significant risk that LASSO may not identify the optimal solution.

In light of the NP-hard nature of the problem, approximation algorithms and heuristic methods are often employed to find satisfactory solutions, even if they are not guaranteed to be optimal. Despite the efforts of researchers like Winker (2000) and Maringer and Winker (2004) in proposing heuristic solutions for model specification in VAR and VECM, the sheer complexity of the problem suggests that identifying the truly optimal model specification remains a formidable challenge.

In our view, the consideration of NP-hard problems receives little to no attention among economists and econometricians in their analyses. Therefore, our paper serves as a prompt and motivation for economists and econometricians to pay more attention and effort to the optimization problem of optimal model specification in the context of NP-hard problems.

This emphasizes the nuanced expertise required in econometrics to adeptly handle VAR and VARX models, striking a delicate balance between model simplicity and the comprehensive capture of economic dynamics. It serves as a reminder that, in econometrics, simplicity often belies underlying complexity, necessitating careful consideration and expertise even in seemingly straightforward analyses.

This challenge becomes notably more daunting when dealing with nonlinear models, a topic discussed in the context of agent-based models by Mandes and Winker (2017). Regarding NLP and LLP modeling, Buveck and Sellke (2021) demonstrate that, generally, larger artificial neural networks with more model parameters yield superior results, advocating for scaling up the size of artificial neural networks. Therefore, the issue of model complexity in empirical research of real-world phenomena remains a frequently underestimated and overlooked challenge.